\documentclass[twocolumn,english]{revtex4-2}
\usepackage[T1]{fontenc}
\usepackage[latin9]{inputenc}
\usepackage{amsmath}
\usepackage{amssymb}
\usepackage{graphicx}
\usepackage{babel}
\begin{document}
\title{Shortcut-to-adiabaticity for coupled harmonic oscillators}
\author{Jonas F. G. Santos}
\email{jonassantos@ufgd.edu.br}

\affiliation{Faculdade de Ci�ncias Exatas e Tecnologia, Universidade Federal da
Grande Dourados, Caixa Postal 364, Dourados, CEP 79804-970, MS, Brazil}
\begin{abstract}
High control in the preparation and manipulation of states is an experimental
and theoretical important task in many quantum protocols. Shortcuts
to adiabaticity methods allow to obtain desirable states of a adiabatic
dynamics but in short time scales. In this work, the problem of considering
this technique for two-coupled bosonic modes is addressed. By using
symplectic transformations it is possible to decouple the modes and
find the driving Hamiltonian that drives the system through a transitionless
dynamics in a finite-time regime. Two paradigmatic interactions are
discussed, the position-position and the magnetic field couplings.
We also illustrate the finding in a concrete example where the nonadiabaticity
degree is quantified in terms of the residual energy. Finally, it
is shown that the driving Hamiltonian for each case contains local
and global contributions, accounting for coherence in the energy basis
and entanglement effects, respectively. 
\end{abstract}
\maketitle

\section{Introduction}

The recent advances in quantum technologies have evidenced the high
control to create and manipulate quantum states \citep{Haase2022,Feng2022,Zheng2022,Cole2021,Sim=0000F3n2020,Xu2023}.
Examples are the preparation of highly entangled states \citep{Li2015}
and Dicke type states \citep{Wang2021}. Despite the vast different
experimental platforms, they all share a common challenge, i.e., the
time requirement to adiabatically prepare quantum states. In general,
the adiabatic preparations demands an effective long time compared
with the time scale of some specific experimental technique. For example,
in nuclear magnetic resonance (NMR), adiabaticity requires processes
evolving on a order of 1ms \citep{Camati2019}, while in some experiments
with trapped ions systems, the adiabaticity conditions is reached
in a time evolution of about 250ms \citep{Cai2021}. Another scenario
in which a short time evolution are desirable is in quantum thermal
machines protocols \citep{Kosloff2014}. Although in many cases the
best efficiency for a given cycle is possible \citep{Pe=0000F1a2020,Cavina2020},
this is only achievable in the adiabatic regime, i.e., the cycle delivers
inevitably a very small power, turning the machine useless in practice.
To circumvent this problem, some authors have employed the efficiency
at maximum power \citep{Curzon1975} in which the focus is the power
in detrimental of the efficiency.

An interesting type of control technique known as shortcuts to adiabaticity
(STA) \citep{Berry2009,Torrontegui2013} have been largely employed
to design protocols, i.e., to prepare quantum states or perform unitary
evolutions, in a desirable time duration. The key idea in STA is to
develop protocols to drive a system by emulating its adiabatic dynamics
on much short time scales. STA protocols are vastly applied to quantum
thermal machines models \citep{Beau2016,Abah2018,Calzetta2018,Hartmann2020}
but not only restricted to them. For instance, STA can be useful in
quantum state preparation protocols \citep{Yuste2013,Abah2020,Chen2021},
quantum computation schemes \citep{Hegade2021}, in systems undergoing
quantum phase transitions \citep{Campbell2015}, as well as to investigate
the Born-Oppenheimer dynamics \citep{Campo2018}. Some experimental
implementations concerning STA protocols have been reported recently
\citep{Vepsalainen2019,Zhou2020}. Moreover, concerning the treatment
of time-dependent harmonic oscillators, it is important to highlight
the Lewis-Riesenfeld (LR) invariant method \citep{Lewis1969}, which
allows to solve the state of the system driven by a known time-dependent
Hamiltonian. In this direction, it has to be mentioned the relevance
of the treatment of time-dependent harmonic oscillators with LR invariant
or other techniques and their applications \citep{Lewis1982,Lewis1967,Dodonov1979,Guedes2001,Moya-Cessa2003,Pedro1997,Fern=0000E1ndez-Guasti2003,Fern=0000E1ndez-Guasti2002,Coelho2022,Patra2023,Onah2023}

Among different systems and scenarios where STA techniques can be
applied, the driving of continuous variables quantum systems is of
particular interest. Again, many quantum thermal machines models employ
single or coupled bosonic modes as working medium, while bosonic modes
can be encoded experimentally in trapped ions systems \citep{Maslennikov2019}.
Entangled states with bosonic modes are also of importance in quantum
information processing. The use of STA in single bosonic modes is
vast, particularly in quantum thermal machines. On the other hand,
for coupled bosonic modes the applications of STA protocols are not
considerably explored. Coupled bosonic modes are relevant in quantum
dynamics, because they can describe, for instance, a chain of harmonic
oscillator transporting energy \citep{Oliveira2017,Iubini2018}, quantum
phase transition systems in the form of the Rabi dimer model \citep{Wang2020},
the interaction of two modes in a nonlinear medium \citep{Mandelbooke}.
Also, there are quantum thermal machines model employing coupled bosonic
as working medium, and a short time scales in these cyclic protocols
is desirable to reach finite power \citep{Sacchi2021}. In Ref. \citep{Campo2018}
the authors considered the STA method for two-coupled harmonic oscillators
for modes with different times scales, where it is possible to decouple
the modes due to the nonidentical dynamics for each mode. 

In this work the problem of finding and discussing the STA for two-mode
coupled bosonic systems are addressed. The issue is initially considered
for general coupling between two bosonic modes in terms of a symplectic
transformation to decouple them. After that, we discuss in detail
two categories of interaction, the position-position and magnetic
field coupling, which represent different physical systems. The discussion
in terms of these couplings is mostly focused on the role played by
the STA driving Hamiltonian, i.e., the local and global part, each
of them avoiding the transition between eigenstates. It is shown that
the local and global part of the STA driving Hamiltonian are associated
with one-mode squeezing (coherence in the energy basis) and two-mode
squeezing (entanglement), respectively. Furthermore, it is discussed
briefly about some physical systems where the STA can be applied for
two-coupled bosonic modes.

The work is organized as follows. Section \ref{sec:Theoretical-framework}
is dedicated to a short review about the STA method for a single harmonic
oscillator. The general method of STA for two-coupled bosonic modes
is also introduced, through a symplectic transformation to decoupled
the modes. In Sec. \ref{sec:Examples} two particular coupling Hamiltonians
are studied in details, and the STA driving Hamiltonian is obtained.
The role played by each part of the STA driving Hamiltonian is discussed.
We also give a concrete example of two-coupled modes with time-dependent
local frequency, and the nonadiabaticity degree is quantified through
the residual energy. The conclusion and final remarks are in Sec.
\ref{sec:Conclusion}.

\section{Theoretical framework \label{sec:Theoretical-framework}}

\subsection{Review on counterdiabatic driving for a single quantum harmonic oscillator}

The central idea of STA techniques is to suppress non-adiabatic transitions
of a quantum systems evolving under a finite-time dynamics, with one
of the main goals being the reduction of the entropy production \citep{Abah2017,Rezek2006}.
Here we review shortly the case for a single quantum harmonic oscillator.
The effective Hamiltonian reads

\begin{equation}
H_{\text{eff}}\left(t\right)=H_{0}\left(t\right)+H_{\text{STA}}\left(t\right).
\end{equation}

Here, $H_{0}\left(t\right)=p^{2}/\left(2m\right)+m\omega^{2}\left(t\right)q^{2}/2$,
with $m$ and $\omega\left(t\right)$ the mass and time-dependent
frequency of the oscillator, and $q$ and $p$ stand for the position
and momentum operators, respectively. Besides, $H_{STA}\left(t\right)$
is STA driving Hamiltonian. By using the counterdiabatic driving,
the aim is to find a Hamiltonian $H_{\text{CD}}\left(t\right)$ for
which the exact solution of the time-dependent Schrodinger equation
for $H_{\text{CD}}\left(t\right)$ matches to the adiabatic approximation
of the original $H_{0}\left(t\right)$. Explicitly, the counterdiabatic
driving reads \citep{Berry2009,Muga2010}

\begin{align}
H_{\text{CD}}\left(t\right) & =H_{0}\left(t\right)+i\hbar\sum_{n}\left(|\partial_{t}n\rangle\langle n|-\langle n|\partial_{t}n\rangle|n\rangle\langle n|\right)\nonumber \\
 & =H_{0}\left(t\right)+H_{\text{STA}}^{\text{CD}}\left(t\right),\label{dd}
\end{align}
with $|n\rangle=|n\left(t\right)\rangle$ the eigenstate of the original
Hamiltonian $H_{0}\left(t\right)$ and $H_{\text{STA}}^{\text{CD}}\left(t\right)$
the STA driving Hamiltonian. In particular, for a single harmonic
oscillator

\begin{align}
H_{\text{STA}}^{\text{CD}}\left(t\right) & =-\frac{\dot{\omega}\left(t\right)}{4\omega\left(t\right)}\left(qp+pq\right)\nonumber \\
 & =i\hbar\frac{\dot{\omega}\left(t\right)}{4\omega\left(t\right)}\left(a^{2}-a^{\dagger2}\right),\label{STAsingle}
\end{align}
 Equation (\ref{dd}) is quadratic in $q$ and $p$, then it may describe
a generalized harmonic oscillator \citep{Muga2010,Chen2010}, i.e.,

\begin{equation}
H_{\text{CD}}\left(t\right)=\frac{p^{2}}{2m}+\frac{m\omega^{2}\left(t\right)}{2}q^{2}-\frac{\dot{\omega}\left(t\right)}{4\omega\left(t\right)}\left(qp+pq\right),
\end{equation}
such that the instantaneous eigenenergies of $H_{\text{CD}}\left(t\right)$
are $E_{n}=\langle H_{\text{CD}}\left(t\right)\rangle=\hbar\omega\left(t\right)Q_{\text{CD}}^{\ast}\left(n+1/2\right)$
\citep{Abah2019}, with $n=1/\left(e^{\beta\hbar\omega\left(0\right)}-1\right)$
the occupation quantum number. The term $Q_{\text{CD}}^{\ast}$ is
called adiabaticity parameter and quantifies how far a time-evolution
is from the adiabatic conditions, with $Q_{\text{CD}}^{\ast}=1$ for
adiabatic dynamics and $Q_{\text{CD}}^{\ast}>1$ otherwise. For a
single harmonic oscillator $Q_{\text{CD}}^{\ast}=1/\sqrt{1-\left(\dot{\omega}\left(t\right)/4\omega^{4}\left(t\right)\right)}$.
Furthermore, the average of the STA control Hamiltonian at any time
is given by \citep{Abah2019}

\begin{equation}
\langle H_{\text{STA}}^{\text{CD}}\left(t\right)\rangle=\frac{\omega\left(t\right)}{\omega\left(0\right)}\left(Q_{\text{CD}}^{\ast}-1\right)\langle H\left(0\right)\rangle,
\end{equation}
with $\langle H\left(0\right)\rangle=\hbar\omega\left(0\right)\coth\left(\beta\hbar\omega\left(0\right)/2\right)/2$.

\subsection{Counterdiabatic driving for coupled harmonic oscillators\label{subsec:Counterdiabatic-driving-for}}

Consider the general coupled Hamiltonian for two quantum harmonic
oscillators

\begin{equation}
H\left(t\right)=H_{0}\left(\boldsymbol{r},t\right)+\Gamma\left(t\right)H_{\text{coup}}\left(\boldsymbol{r},t\right),
\end{equation}
where $\boldsymbol{r=}\boldsymbol{r}\left(q_{1},p_{1},q_{2},p_{2}\right)$
is now the quadrature vector and $\Gamma$ is the coupling strength.
Assuming that $H\left(t\right)$ is at most quadratic in position
and momentum, it is possible to find a symplectic transformation $S\left(\boldsymbol{r}\right)$
to write the Hamiltonian $H\left(t\right)$ in terms of their normal
modes \citep{SerafiniBooks}, i.e.,

\begin{equation}
S\left(\boldsymbol{r}\right)H\left(t\right)S\left(\boldsymbol{r}\right)^{\intercal}\rightarrow\mathcal{H}\left(\boldsymbol{R}\right)=\mathcal{H}_{1}\left(\boldsymbol{R_{1}}\right)+\mathcal{H}_{2}\left(\boldsymbol{R_{2}}\right),
\end{equation}
with $\boldsymbol{R}=\boldsymbol{R}\left(Q_{1},P_{1},Q_{2},P_{2}\right)$
the normal modes quadrature vector, $\boldsymbol{R}_{1}=\boldsymbol{R}_{1}\left(Q_{1},P_{1}\right)$
, $\boldsymbol{R}_{2}=\boldsymbol{R}_{2}\left(Q_{2},P_{2}\right)$,
and $S\left(\boldsymbol{r}\right)$ depends on the explicit form of
$H_{\text{coup}}\left(\boldsymbol{r},t\right)$ . Since $\mathcal{H}\left(\boldsymbol{R}\right)$
is effectively two uncoupled quantum harmonic oscillators, we can
apply the counterdiabatic driving for each mode, resulting in

\begin{equation}
\mathcal{H}_{i}^{CD}\left(t\right)=\mathcal{H}_{i}\left(t\right)+\mathcal{H}_{STA,i}\left(t\right),
\end{equation}
 with $i=1,2$. This implies that

\begin{equation}
\mathcal{H}^{CD}\left(t\right)=\mathcal{H}\left(t\right)+\mathcal{H}_{STA}\left(t\right),
\end{equation}
where $\mathcal{H}\left(t\right)=\mathcal{H}_{1}\left(t\right)+\mathcal{H}_{2}\left(t\right)$
and $\mathcal{H}_{STA}\left(t\right)=\mathcal{H}_{STA,1}\left(t\right)+\mathcal{H}_{STA,2}\left(t\right)$.
Note that for the normal modes counterdiabatic Hamiltonian the frequency
will in general depend on the coupling strength. Below we explore
two relevant interactions to illustrate this result.

\section{Discussions on two interactions\label{sec:Examples}}

\subsection*{Position-position coupling}

In order to discuss the method for relevant interactions, we first
consider the two-coupled harmonic oscillators with a position-position
coupling. The Hamiltonian can be written as 

\begin{equation}
H\left(t\right)=\sum_{i=1}^{2}\left(\frac{p_{i}^{2}}{2m}+\frac{m\omega^{2}\left(t\right)q_{i}^{2}}{2}\right)+\gamma\left(t\right)q_{1}q_{2},\label{Hqq}
\end{equation}
where $m$ and $\omega\left(t\right)$ are the mass and time-dependent
frequency of the free oscillators, respectively and $\gamma\left(t\right)$
is the time-dependent coupling strength. Examples of the use of this
interaction are the energy transport in a chain of quantum harmonic
oscillators \citep{Oliveira2017,Iubini2018} as well as the lowest
spin energy projection of the Rabi dimer model \citep{Wang2020}.
The Hamiltonian in Eq. (\ref{Hqq}) is general because is encompasses
time-dependence in the frequency of the local oscillators as well
as in the coupling between them. Let assume that the initial state
is written as $\rho_{0}=\rho_{0}^{1}\otimes\rho_{0}^{2}$. By imposing
the STA method one has

\begin{equation}
H^{CD}\left(t\right)=H\left(t\right)+H_{STA}^{CD}\left(t\right).\label{HqqCD}
\end{equation}

In order to find the mathematical expression of $H_{STA}^{CD}\left(t\right)$,
we apply the following variable transformation on $H\left(t\right)$

\begin{equation}
\left(\begin{array}{c}
q_{1}\\
q_{2}\\
p_{1}\\
p_{2}
\end{array}\right)=\left(\begin{array}{cccc}
1 & 1 & 0 & 0\\
1 & -1 & 0 & 0\\
0 & 0 & 1 & 1\\
0 & 0 & 1 & -1
\end{array}\right)\left(\begin{array}{c}
Q_{1}\\
Q_{2}\\
P_{1}\\
P_{2}
\end{array}\right),
\end{equation}

which results in the following Hamiltonian

\begin{align}
\mathcal{H}\left(t\right) & =\sum_{i=1}^{2}\left(\frac{P_{i}^{2}}{2m}+\frac{m\omega_{i}^{2}\left(t\right)}{2}Q_{i}^{2}\right)\\
 & =\mathcal{H}_{1}^{0}\left(t\right)+\mathcal{H}_{2}^{0}\left(t\right)
\end{align}
with $\omega_{1}^{2}\left(t\right)=\omega^{2}\left(t\right)+\gamma\left(t\right)/m$
and $\omega_{2}^{2}\left(t\right)=\omega^{2}\left(t\right)-\gamma\left(t\right)/m$.
From section \ref{subsec:Counterdiabatic-driving-for} we can write
for each local harmonic oscillator above

\begin{align}
\mathcal{H}_{1}^{0}\left(t\right) & =\frac{P_{1}^{2}}{2m}+\frac{m\omega_{1}^{2}\left(t\right)}{2}Q_{1}^{2},\\
\mathcal{H}_{2}^{0}\left(t\right) & =\frac{P_{2}^{2}}{2m}+\frac{m\omega_{2}^{2}\left(t\right)}{2}Q_{2}^{2}.\label{extra03}
\end{align}

For these two local Hamiltonians with different frequencies, one locally
applies the STA method, such that

\begin{align}
\mathcal{H}_{1}^{CD}\left(t\right) & =\mathcal{H}_{1}^{0}\left(t\right)+\mathcal{H}_{STA,1}^{CD}\left(t\right)\nonumber \\
\mathcal{H}_{2}^{CD}\left(t\right) & =\mathcal{H}_{2}^{0}\left(t\right)+\mathcal{H}_{STA,2}^{CD}\left(t\right),\label{new01}
\end{align}

with

\begin{align}
\mathcal{H}_{STA,1}^{CD}\left(t\right) & =-\frac{\dot{\omega}_{1}}{4\omega_{1}}\left(Q_{1}P_{1}+P_{1}Q_{1}\right),\label{extra01}\\
\mathcal{H}_{STA,2}^{CD}\left(t\right) & =-\frac{\dot{\omega}_{2}}{4\omega_{2}}\left(Q_{2}P_{2}+P_{2}Q_{2}\right).\label{extra02}
\end{align}
 By applying the inverse transformation to obtain $H_{STA,}^{CD}\left(t\right)$
in Eq. (\ref{HqqCD}) one gets

\begin{align}
H_{STA,}^{CD}\left(t\right) & =-F\left(t\right)\left(q_{1}p_{1}+p_{1}q_{1}+q_{2}p_{2}+p_{2}q_{2}\right)\nonumber \\
 & -G\left(t\right)\left(q_{1}p_{2}+q_{2}p_{1}\right),\label{counterexample01}
\end{align}
with

\begin{align}
F\left(t\right) & =\left(\frac{\dot{\omega}_{1}\left(t\right)}{8\omega_{1}\left(t\right)}+\frac{\dot{\omega}_{2}\left(t\right)}{8\omega_{2}\left(t\right)}\right)\nonumber \\
G\left(t\right) & =\left(\frac{\dot{\omega}_{1}\left(t\right)}{4\omega_{1}\left(t\right)}-\frac{\dot{\omega}_{2}\left(t\right)}{4\omega_{2}\left(t\right)}\right).\label{new03}
\end{align}

This counterdiabatic driving Hamiltonian shows explicitly the role
played by the coupling. It is instructive to consider some particular
examples. First, if we assume $\gamma\left(t\right)=0$, then $\dot{\omega}_{1}^{2}(t)=\dot{\omega}_{2}^{2}(t)=\dot{\omega}^{2}(t)$
and

\begin{equation}
H_{STA}^{CD}\left(t\right)=-\frac{1}{4}\left(\frac{\dot{\omega}\left(t\right)}{\omega\left(t\right)}\right)\left(q_{1}p_{1}+p_{1}q_{1}+q_{2}p_{2}+p_{2}q_{2}\right),\label{first}
\end{equation}
 which is the standard case for two-uncoupled harmonic oscillators
\citep{Muga2010}. The second case is when $\gamma(t)=\gamma$, i.e.,
a time-independent coupling. In this case

\begin{align}
\dot{\omega}_{1\left(2\right)}\left(t\right) & =\frac{\omega(t)\dot{\omega}(t)}{\sqrt{\omega^{2}(t)\pm\gamma(t)/m}},
\end{align}
 resulting in

\begin{align}
H_{STA}^{CD}\left(t\right) & =-\frac{1}{4}\frac{\omega^{3}\left(t\right)\dot{\omega}\left(t\right)}{\left(\omega^{4}\left(t\right)-\gamma^{2}/m^{2}\right)}\left(q_{1}p_{1}+p_{1}q_{1}+q_{2}p_{2}+p_{2}q_{2}\right)\nonumber \\
 & +\frac{\omega\left(t\right)\dot{\omega}\left(t\right)\gamma}{2\left(\omega^{4}\left(t\right)-\gamma^{2}/m^{2}\right)}\left(q_{1}p_{2}+q_{2}p_{1}\right).\label{second}
\end{align}

Finally, the third case is when $\omega\left(t\right)=\omega$ and
the time-dependence is only on $\gamma\left(t\right)$. In this case

\begin{align}
H_{STA,}^{CD}(t) & =\frac{\gamma\left(t\right)\dot{\gamma}\left(t\right)}{8\left(m^{2}\omega^{4}-\gamma^{2}\left(t\right)\right)}\left(q_{1}p_{1}+p_{1}q_{1}+q_{2}p_{2}+p_{2}q_{2}\right)\nonumber \\
 & -\frac{m\omega^{2}\dot{\gamma}\left(t\right)}{8\left(m^{2}\omega^{4}-\gamma^{2}\left(t\right)\right)}\left(q_{1}p_{2}+q_{2}p_{1}\right).\label{third}
\end{align}

The physical meaning of the counterdiabatic driving Hamiltonian in
Eq. (\ref{counterexample01}) can be understood using standard bosonic
annihilation and creation operators such that $a_{1\left(2\right)}|n_{1\left(2\right)}\rangle=\sqrt{n_{1\left(2\right)}}|n_{1\left(2\right)}-1\rangle$
and $a_{1\left(2\right)}^{\dagger}|n_{1\left(2\right)}\rangle=\sqrt{n_{1\left(2\right)}+1}|n_{1\left(2\right)}+1\rangle$.
It is shown that $\left(q_{i}p_{i}+p_{i}q_{i}\right)\propto\left(a_{i,t}^{2}-a_{i,t}^{\dagger2}\right)$,
while $\left(q_{1}p_{2}+q_{2}p_{1}\right)\propto\left(a_{1,t}a_{2,t}-a_{1,t}^{\dagger}a_{2,t}^{\dagger}\right)$,
evidencing that the local part of Eq. (\ref{counterexample01}) is
responsible for one-mode squeezing, while the coupling part is responsible
for two-mode squeezing. Thus, since a finite-time dynamics of a two-coupled
quantum harmonic oscillators generates correlations between the modes,
the two-mode squeezing effect in Eq. (\ref{counterexample01}) is
responsible for maintaining the dynamics transitionless in a global
level.

We now turn on our attention to the adiabaticity parameter $Q^{\ast}$
in this example using Eq. (\ref{new01}). Both $\mathcal{H}_{1}\left(t\right)$
and $\mathcal{H}_{2}^{CD}\left(t\right)$ are quadratic in position
and momentum, so that they can be described in terms of a generalized
harmonic oscillator with non-local operator \citep{Muga2010,Chen2010},

\[
\mathcal{H}_{i}^{CD}\left(t\right)=\frac{P_{i}^{2}}{2m}+\frac{m\omega_{i}^{2}\left(t\right)}{2}Q_{i}^{2}-\frac{\dot{\omega}_{i}}{4\omega_{i}}\left(Q_{i}P_{i}+P_{i}Q_{i}\right),
\]
with the instantaneous eigenenergies for each mode given by $E_{n}=\langle\mathcal{H}_{i}^{CD}\left(t\right)\rangle=\hbar\omega_{i}\left(t\right)Q_{CD,i}^{\ast}\left(\bar{n}_{i}+1/2\right)$,
with $\bar{n}_{i}$ the average excitation number for each mode and
$Q_{CD,i}^{\ast}=1/\sqrt{1-\left(\dot{\omega}_{i}\left(t\right)/4\omega_{i}^{4}\left(t\right)\right)}$.
Furthermore, the average of the control $\langle H_{STA}^{CD}\left(t\right)\rangle=\langle\mathcal{H}_{STA}^{CD}\left(t\right)\rangle$
is given

\[
\langle H_{STA}^{CD}\left(t\right)\rangle=\sum_{i=1}^{2}\frac{\omega_{i}\left(t\right)}{\omega_{i}\left(0\right)}\left(Q_{CD,i}^{\ast}-1\right)\langle\mathcal{H}_{i}^{0}\left(0\right)\rangle,
\]
where $\langle\mathcal{H}_{i}^{0}\left(0\right)\rangle=\left(\hbar\omega_{i}\left(0\right)/2\right)\coth\left(\beta_{i}\hbar\omega_{i}\left(0\right)/2\right).$
This last result is important because it shows that the adiabatic
parameter for coupled quantum oscillators depends not only on the
shape of the driving but also on how the coupling strength is modulated
during the dynamics. Typical applications of this coupling structure
are when the interaction is set to be \citep{Souza2022,Joshi2014}

\begin{equation}
U=\lambda\left(a^{\dagger}b+ab^{\dagger}\right)+\mu\left(ab+a^{\dagger}b^{\dagger}\right),
\end{equation}
where $\lambda$ and $\mu$ tune the strength of the exchange of excitation
and two-mode squeezing, respectively, such that $\mu\ll\omega$ implies
the squeezing coupling is weak and can be neglected, implying the
validity of the rotate wave approximation (RWA), while for $\mu\sim\omega$
the squeezing coupling can not be discard, and the RWA is no longer
valid. For this application, $\gamma\left(t\right)$ in Eq. (\ref{Hqq})
is a function of $\lambda$ and/or $\mu$, depending on which regime
is considered. The method is also applicable to systems describing
quantum phase transition, for instance, in finite-component systems,
like the the quantum Rabi dimer, with interaction Hamiltonian given
by \citep{Wang2020,Mao2021}

\begin{equation}
J\left(a+a^{\dagger}\right)\left(b+b^{\dagger}\right),
\end{equation}
 in which $\gamma\left(t\right)=J$ in this case. The particularity
of the quantum Rabi dimer is that there is no a regime where the RWA
applies. Here the STA method for two-modes could be relevant to suppress
correlations arising near the critical point.

\subsection*{Magnetic field interaction}

As a second relevant form for the coupling, we consider that representing
the two quantum harmonic oscillator interacting through a magnetic
field with general frequency $\omega_{B}\left(t\right)$. The Hamiltonian
in this case is given by

\begin{equation}
H=\alpha^{2}p_{i}^{2}+\beta^{2}(t)q_{i}^{2}+\omega_{B}(t)\sum_{i,j=1}^{2}p_{i}q_{j},\label{secondcase}
\end{equation}
with $\alpha^{2}=1/2m$, $\beta^{2}(t)=\frac{m}{2}\Omega^{2}(t)$,
and $\Omega^{2}(t)=\omega_{0}^{2}+\omega_{B}^{2}(t)$. Writing this
Hamiltonian in terms of $a_{i}\left(a_{i}^{\dagger}\right)$ one has

\begin{equation}
H=\hbar\Omega(t)\left[\left(a_{1}^{\dagger}a_{1}+a_{2}^{\dagger}a_{2}+1\right)\right]-i\omega_{B}(t)\hbar\left[a_{1}a_{2}^{\dagger}-a_{1}^{\dagger}a_{2}\right],
\end{equation}
where we highlight the use of such a coupling in thermodynamics cycles
of Gaussian bipartite states \citep{Sacchi2021}. Moreover, the time
evolution using this coupling is important in quantum optics, considering
the interaction of two modes in a nonlinear medium \citep{Mandelbooke}.
These two examples show that finding a STA method is relevant to obtain
desirable final two-mode states.

By performing the following variable transformation on Eq.(\ref{secondcase})

\begin{equation}
\left(\begin{array}{c}
Q_{+}\\
P_{+}\\
Q_{-}\\
P_{-}
\end{array}\right)=\left(\begin{array}{cccc}
\delta_{+}\beta & 0 & 0 & \delta_{+}\alpha\\
0 & \epsilon_{+} & -\epsilon_{+}\beta & 0\\
\delta_{-}\beta & 0 & 0 & -\delta_{-}\alpha\\
0 & \epsilon_{-}\alpha & \epsilon_{-}\beta & 0
\end{array}\right)\left(\begin{array}{c}
q_{1}\\
p_{1}\\
q_{2}\\
p_{2}
\end{array}\right),
\end{equation}
with $\delta_{\pm}=\sqrt{1/m\omega_{\pm}\Omega}$, $\epsilon_{\pm}=\sqrt{m\omega_{\pm}/\Omega}$,
$\omega_{\pm}=\Omega\mp\omega_{B}$, $\Omega^{2}=\omega_{0}^{2}+\omega_{B}^{2}$,
and omitting the time-dependence in the quantities for a compact notation,
one has

\[
\mathcal{H}\left(t\right)=\mathcal{H}_{+}^{\left(0\right)}\left(t\right)+\mathcal{H}_{-}^{\left(0\right)}\left(t\right),
\]
where

\begin{equation}
\mathcal{H}_{\pm}^{\left(0\right)}\left(t\right)=\frac{\mathcal{P}_{\pm}^{2}}{2m}+\frac{m\omega_{\pm}^{2}}{2}\mathcal{Q}_{\pm}^{2}.
\end{equation}

Applying the STA method here, we have the following

\[
\mathcal{H}_{STA,\pm}^{CD}\left(t\right)=\mathcal{H}_{\pm}^{\left(0\right)}\left(t\right)+\mathcal{H}_{STA,\pm}^{CD}\left(t\right),
\]
with

\begin{equation}
\mathcal{H}_{STA,\pm}^{CD}\left(t\right)=-\frac{\dot{\omega}_{\pm}}{4\omega_{\pm}}\left(Q_{\pm}P_{\pm}+P_{\pm}Q_{\pm}\right),
\end{equation}
 and the inverse transformation results in

\begin{align}
H_{STA}^{CD}\left(t\right) & =-M\left(t\right)\left(q_{1}p_{1}+p_{1}q_{1}-q_{2}p_{2}-p_{2}q_{2}\right)\nonumber \\
 & -N\left(t\right)\left(\frac{\alpha}{\beta\left(t\right)}p_{1}p_{2}-\frac{\beta\left(t\right)}{\alpha}q_{1}q_{2}\right),\label{STAsecond}
\end{align}
with

\begin{align*}
M\left(t\right) & =\frac{\dot{\Omega}\left(t\right)\Omega\left(t\right)-\dot{\omega}_{B}\left(t\right)\omega_{B}\left(t\right)}{4\omega_{0}^{2}\left(t\right)},\\
N\left(t\right) & =2\frac{\dot{\Omega}\left(t\right)\omega_{B}-\dot{\omega}_{B}\Omega}{\omega_{0}^{2}\left(t\right)}.
\end{align*}

Again, some particular cases are relevant here. Firstly, for $\omega_{B}\left(t\right)=0$
one gets

\begin{equation}
H_{STA}^{CD}\left(t\right)=-\frac{\dot{\omega_{0}}\left(t\right)}{4\omega_{0}\left(t\right)}\left(q_{1}p_{1}+p_{1}q_{1}-q_{2}p_{2}-p_{2}q_{2}\right),
\end{equation}
which is exactly the counterdiabatic driving Hamiltonian for two-uncoupled
orthogonal quantum harmonic oscillators. The second relevant case
is $\omega_{B}\left(t\right)=\omega_{B}$, i.e., a constant coupling,
which results in

\begin{align}
H_{STA}^{CD}\left(t\right) & =-\frac{\dot{\omega_{0}}\left(t\right)\Omega\left(t\right)}{2\omega_{0}\left(t\right)}\left(q_{1}p_{1}+p_{1}q_{1}-q_{2}p_{2}-p_{2}q_{2}\right)\nonumber \\
 & -\frac{4\dot{\omega}_{0}\left(t\right)\omega_{B}}{\omega_{0}\left(t\right)}\left(\frac{\alpha}{\beta\left(t\right)}p_{1}p_{2}-\frac{\beta\left(t\right)}{\alpha}q_{1}q_{2}\right).\label{secondcasesecond}
\end{align}

Here, we also note that the coupling between the two modes implies
a term corresponding to global correlations in the counterdiabatic
driving Hamiltonian, Eqs. (\ref{STAsecond}) and (\ref{secondcasesecond}).
To be more explicit, note that both $q_{1}q_{2}$ and $p_{1}p_{2}$
are proportional to $a_{1}^{\dagger}a_{2}^{\dagger}+a_{1}a_{2}\pm a_{1}^{\dagger}a_{2}\pm a_{1}a_{2}^{\dagger}$,
with the sign $\pm$ depending on what product we are considering.
Thus, we can conclude that the magnetic field coupling in the two-mode
Hamiltonian generates a correlation term in the counterdiabatic driving
Hamiltonian that is responsible for two-mode squeezing and excitation
change between the modes \citep{Souza2022}

A possible application of the method for magnetic field coupling is
in quantum thermal machines \citep{Sacchi2021}, where two-coupled
harmonic oscillators are considered as the working substance. In that
case the interaction strength is associated to the nonlinear medium
property and it is equivalent to a unitary transformation with coupling
$\xi=\theta e^{i\varphi}$, with $\theta\left(t\right)$ representing
the time-dependence of the coupling. Since this engine model may create
entanglement due to the nature of the interaction \citep{Reb=0000F3n2011},
the STA method for two modes could be employed to suppress this quantum
correlation in a arbitrary finite-time regime, impacting its performance.

\subsection*{Example}

In the following, we consider an example where two quantum harmonic
oscillators are coupled through position-position coupling. The Hamiltonian
is

\begin{align}
H\left(t\right) & =\sum_{i=1}^{2}\left[\omega_{0}\left(1-\frac{g^{2}}{2}\right)a_{i}^{\dagger}a_{i}-\omega_{0}\frac{g^{2}}{4}\left(a_{i}^{\dagger2}+a_{i}^{2}+1\right)\right]\nonumber \\
 & +\omega_{0}\frac{J}{2}\left(a_{1}^{\dagger}+a_{1}\right)\left(a_{2}^{\dagger}+a_{2}\right),\label{ex01}
\end{align}
where $g=g\left(t\right)$ introduces a time-dependent behavior in
the local systems and $J=J\left(t\right)$ mediates the coupling between
them. Physically, this Hamiltonian represents two Rabi models interacting
through cavity modes when the spins are projected onto the lowest
eigenstates \citep{Wang2020,Mao2021}. To decouple the modes, we apply
the transformation

\begin{align}
c & =\frac{1}{\sqrt{2}}\left(a_{1}+a_{2}\right),\nonumber \\
d & =\frac{1}{\sqrt{2}}\left(a_{1}-a_{2}\right),\label{ex02}
\end{align}
with $\left[c,c^{\dagger}\right]=\left[d,d^{\dagger}\right]=1$. After
some manipulation, the complete Hamiltonian reads

\begin{equation}
H\left(t\right)=\mathcal{H}_{1}\left(t\right)+\mathcal{H}_{2}\left(t\right),\label{ex03}
\end{equation}
where

\begin{align*}
\mathcal{H}_{1}\left(t\right) & =\omega_{1}\left(t\right)c^{\dagger}c+\frac{\omega_{0}}{4}g_{1}^{2}\left(c^{\dagger2}+c^{2}\right)-\omega_{0}\frac{g^{2}}{4},\\
\mathcal{H}_{2}\left(t\right) & =\omega_{2}\left(t\right)d^{\dagger}d-\frac{\omega_{0}}{4}g_{2}^{2}\left(d^{\dagger2}+d^{2}\right)-\omega_{0}\frac{g^{2}}{4},
\end{align*}
 and with the definitions $\omega_{1}\left(t\right)=\omega_{0}\left(1+\frac{1}{2}\left(J-g^{2}\right)\right)$
and $\omega_{2}\left(t\right)=\omega_{0}\left(1-\frac{1}{2}\left(J+g^{2}\right)\right)$,
$g_{1}=\sqrt{g^{2}-J}$, $g_{2}=\sqrt{g^{2}+J}$. The Hamiltonians
$\mathcal{H}_{1}\left(t\right)$ and $\mathcal{H}_{2}\left(t\right)$
can be diagonalized to the standard form in Eq. (\ref{extra03}) (See
appendix).

We now consider the insertion of the STA driving Hamiltonian. In Eq.
(\ref{STAsingle}) the STA driving Hamiltonian is expressed for a
unsqueezed quantum harmonic oscillator. However, for the present example,
the diagonalization of $\mathcal{H}_{1}\left(t\right)$ and $\mathcal{H}_{2}\left(t\right)$
leads to squeezed quantum harmonic oscillators. 

In Appendix, we show that the mathematical form for the STA driving
Hamiltonian is exactly the same for both cases, squeezed and unsqueezed.
Following Eqs. (\ref{extra01}) and (\ref{extra02}) for the decoupled
modes or analogously Eq. (\ref{counterexample01}) for the original
modes, we find

\begin{align*}
\mathcal{H}_{STA,1}^{CD}\left(t\right) & =-\frac{\dot{\omega}_{1}}{4\omega_{1}}\left(c^{2}-c^{\dagger2}\right),\\
\mathcal{H}_{STA,2}^{CD}\left(t\right) & =-\frac{\dot{\omega}_{2}}{4\omega_{2}}\left(d^{2}-d^{\dagger2}\right),
\end{align*}
and 

\begin{align*}
\mathcal{H}_{1}^{CD}\left(t\right) & =\mathcal{H}_{1}\left(t\right)+\mathcal{H}_{STA,1}^{CD}\left(t\right),\\
\mathcal{H}_{2}^{CD}\left(t\right) & =\mathcal{H}_{2}\left(t\right)+\mathcal{H}_{STA,2}^{CD}\left(t\right).
\end{align*}

In order to consider the dynamics of the model, we move to the Heisenberg
picture and write the time evolution of the operators $c$ and $d$
in the form

\begin{align}
c_{H}\left(t\right) & =u_{1}\left(t\right)c+v_{1}^{\ast}\left(t\right)c^{\dagger},\nonumber \\
d_{H}\left(t\right) & =u_{2}\left(t\right)d+v_{2}^{\ast}\left(t\right)d^{\dagger},\label{ex04}
\end{align}
 where the subscript $H$ indicates the Heisenberg picture and the
time-dependent functions satisfy the constraint $|u_{i}|^{2}-|v_{i}|^{2}=1$,
with $i=1,2$. Using the Heisenberg equation of motion, $i\dot{c}_{H}\left(t\right)=\left[c_{H}\left(t\right),\mathcal{H}_{1}^{CD}\left(t\right)\right]$
and $i\dot{d}_{H}\left(t\right)=\left[d_{H}\left(t\right),\mathcal{H}_{2}^{CD}\left(t\right)\right]$,
we obtain the following coupled differential equations

\begin{align}
i\frac{du_{1}}{dt} & =\omega_{0}\left[\left(1-\frac{g_{1}^{2}}{2}\right)u_{1}-\frac{g_{1}^{2}}{2}v_{1}\right]-i\frac{\dot{\varpi}_{t,1}}{2\omega_{t,1}}v_{1},\nonumber \\
-i\frac{dv_{1}}{dt} & =\omega_{0}\left[\left(1-\frac{g_{1}^{2}}{2}\right)v_{1}-\frac{g_{1}^{2}}{2}u_{1}\right]+i\frac{\dot{\varpi}_{t,1}}{2\omega_{t,1}}u_{1},\nonumber \\
i\frac{du_{2}}{dt} & =\omega_{0}\left[\left(1-\frac{g_{2}^{2}}{2}\right)u_{2}-\frac{g_{2}^{2}}{2}v_{2}\right]-i\frac{\dot{\varpi}_{t,2}}{2\omega_{t,2}}v_{2},\nonumber \\
-i\frac{dv_{2}}{dt} & =\omega_{0}\left[\left(1-\frac{g_{2}^{2}}{2}\right)v_{2}-\frac{g_{2}^{2}}{2}u_{2}\right]+i\frac{\dot{\varpi}_{t,2}}{2\varpi_{t,2}}u_{2},\label{ex05}
\end{align}
where we defined $\varpi_{t,1}=\sqrt{\omega_{0}^{2}\left(1-g_{1}^{2}\left(t\right)\right)}$
and $\varpi_{t,2}=\sqrt{\omega_{0}^{2}\left(1-g_{2}^{2}\left(t\right)\right)}$.

In order to compute how the STA method works on the dynamics of two-coupled
quantum harmonic oscillator, we consider the residual energy $E_{r}$,
which measures the degree of nonadiabaticity of the dynamics and it
is defined, for our example, by

\[
E_{r}=\langle0|\mathcal{H}_{1}^{CD}\left(t\right)|0\rangle+\langle0|\mathcal{H}_{2}^{CD}\left(t\right)|0\rangle-E_{G_{1}}-E_{G_{2}},
\]
with 

\begin{align*}
\langle0|\mathcal{H}_{1}^{CD}\left(t\right)|0\rangle & =\omega_{0}|v_{1}|^{2}-\frac{\omega_{0}g_{1}^{2}}{4}|u_{1}+v_{1}|^{2}-\frac{\omega_{0}}{4}J\\
 & +i\frac{\dot{\varpi}_{t,1}}{4\varpi_{t,1}}\left(u_{1}\left(\tau_{q}\right)v_{1}^{\ast}\left(\tau_{q}\right)-u_{1}^{\ast}\left(\tau_{q}\right)v_{1}\left(\tau_{q}\right)\right),\\
\langle0|\mathcal{H}_{2}^{CD}\left(t\right)|0\rangle & =\omega_{0}|v_{2}|^{2}-\frac{\omega_{0}g_{2}^{2}}{4}|u_{2}+v_{2}|^{2}+\frac{\omega_{0}}{4}J\\
 & +i\frac{\dot{\varpi}_{t,2}}{4\varpi_{t,2}}\left(u_{2}\left(\tau_{q}\right)v_{2}^{\ast}\left(\tau_{q}\right)-u_{2}^{\ast}\left(\tau_{q}\right)v_{2}\left(\tau_{q}\right)\right),
\end{align*}
and we are assuming the system prepared in the ground state. We reinforce
that the quantity $E_{G}=E_{G_{1}}+E_{G_{2}}$ is the ground state
energy if the dynamics is adiabatic. 

\begin{figure}
\includegraphics[scale=0.55]{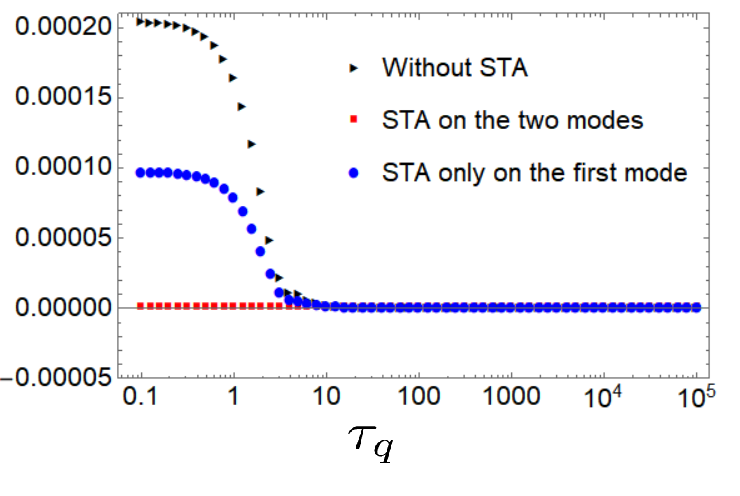}

\includegraphics[scale=0.53]{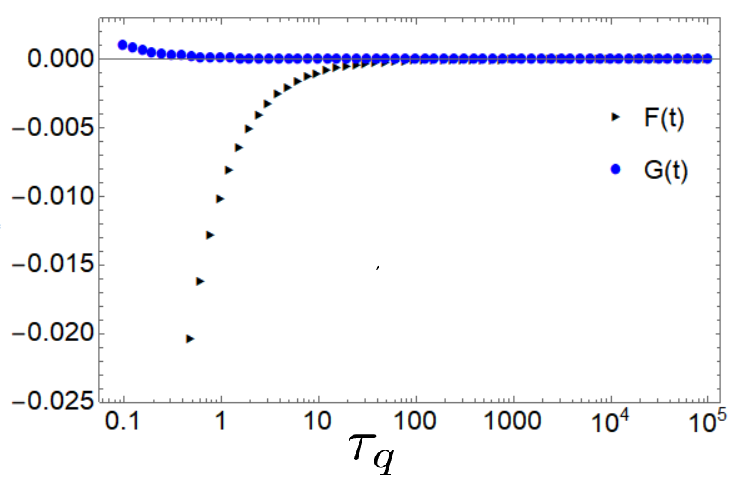}\caption{Dynamics. (Top figure): Residual energy as a function of time for
two-coupled harmonic oscillator in which $g\left(t\right)=g_{f}t/\tau_{q}$
and $J$ is a constant. (Bottom figure): Time evolution of $F\left(t\right)$
and $G\left(t\right)$, defined in Eq. (\ref{new03}), as a function
of time for the present example. The behavior is exactly the same
for the three cases considered in the Top Figure. We considered $\omega_{0}=1$,
$g_{f}=0.2$, $J=0.01$. }

\label{Exampe01}
\end{figure}

We consider the case in which $g\left(t\right)=g_{f}t/\tau_{q}$ and
$J$ is a constant, with $\tau_{q}$ a time-scale of the dynamics.
 \ref{Exampe01} (Top Figure) shows the residual energy $E_{r}$
as a function of time for three cases, for both modes without STA
driving (black markers), with STA driving applied for only the first
mode (blue markers), and for STA applied for both modes (red markers).
It is possible to observe that when the STA driving Hamiltonian is
applied on both modes, the dynamics of the system is very close to
the adiabatic one, even for short times. Although the present example
with $\omega=\omega\left(t\right)$ and $J$ constant is more relevant,
the method could also be applied if $J=J\left(t\right)$. Furthermore,
Fig. \ref{Exampe01} (Bottom Figure) depicts the functions $F\left(t\right)$
and $G\left(t\right)$, defined in Eq. (\ref{new03}), as a function
of time for the present example. Based on the behavior of $F\left(t\right)$
and $G\left(t\right)$ it is possible to have the time-scale in which
the dynamics becomes adiabatic, around $\text{\ensuremath{\tau_{q}}}\sim100$
for the set of parameters considered. This time-scale agrees, as expect,
with the dynamics of the residual energy. Moreover, we can also conclude
that, for the present example, the local squeezing effect is greater
than the global correlations.

\section{Conclusion\label{sec:Conclusion}}

In this work we extended the counterdiabatic driving Hamiltonian method,
based on the STA strategies, for two-coupled harmonic oscillators.
The general formalism consists in using symplectic transformation
to decouple the system and then applying one-mode STA Hamiltonian
for each normal mode. The symplectic transformation depends on the
structure of the coupling Hamiltonian. We have considered two relevant
interactions to discuss in detail, the position-position and magnetic
field couplings, as well as a complete example where the nonadiabaticity
degree is quantified in terms of the residual energy.

Our results elucidates that, while one-mode STA Hamiltonians aim to
control local quantum coherence and one-mode squeezing, the two-mode
STA Hamiltonian controls global correlations, as entanglement and
two-mode squeezing. We also highlighted some physical situations where
this method could be applied, as in the development of finite-time
thermal machines fueled by two-mode harmonic oscillators and finite-component
systems presenting quantum phase transitions. 
\begin{acknowledgments}
We acknowledge Universidade Federal da Grande Dourados, CAPES and
CNPq Grant No. 420549/2023-4 for the support. \\

\textbf{Data Availability Statement:} No data associated in the manuscript.
\end{acknowledgments}

\section*{Appendix}

\subsection*{Diagonalization}

Here we start considering the Hamiltonians 

\begin{align}
\mathcal{H}_{1}\left(t\right) & =\omega_{1}\left(t\right)c^{\dagger}c+\frac{\omega_{0}}{4}g_{1}^{2}\left(c^{\dagger2}+c^{2}\right)-\omega_{0}\frac{g^{2}}{4},\nonumber \\
\mathcal{H}_{2}\left(t\right) & =\omega_{2}\left(t\right)d^{\dagger}d-\frac{\omega_{0}}{4}g_{2}^{2}\left(d^{\dagger2}+d^{2}\right)-\omega_{0}\frac{g^{2}}{4},\label{A01}
\end{align}
 which can be written in the form

\begin{align}
\mathcal{H}_{1} & =\omega_{0}c^{\dagger}c-\frac{\omega_{0}}{4}g_{1}^{2}\left(c^{\dagger}+c\right)^{2}-\frac{\omega_{0}}{4}J,\nonumber \\
\mathcal{H}_{2} & =\omega_{0}d^{\dagger}d-\frac{\omega_{0}}{4}g_{2}^{2}\left(d^{\dagger}+d\right)^{2}+\frac{\omega_{0}}{4}J,\label{A02}
\end{align}
where we used the relation $c^{\dagger2}+c^{2}=\left(c^{\dagger}+c\right)^{2}-2c^{\dagger}c-1$,
and $d^{\dagger2}+d^{2}=\left(d^{\dagger}+d\right)^{2}-2d^{\dagger}d-1$.
For the mode $c$ (and analogously for the mode $d$) we apply the
squeezing operator $S\left[r_{c}\right]=\exp\left[\frac{r_{c}}{2}\left(c^{\dagger2}-c^{2}\right)\right]$
such that the result is

\begin{align*}
S^{\dagger}\left[r_{c}\right]\mathcal{H}_{1}S\left[r_{c}\right] & =\omega_{0}\left\{ \cosh\left(2r_{c}\right)-\frac{g_{1}^{2}}{2}e^{2r_{c}}\right\} \left(c^{\dagger}c+\frac{1}{2}\right)\\
 & +\frac{\omega_{0}}{2}\left\{ \sinh\left(2r_{c}\right)-\frac{g_{1}^{2}}{2}e^{2r_{c}}\right\} \left(c^{2}+c^{\dagger2}\right)\\
 & -\frac{\omega_{0}}{2}-\frac{\omega_{0}}{4}J,
\end{align*}
and by requiring that the term $\left(c^{2}+c^{\dagger2}\right)$
vanishes, we find

\begin{align*}
r_{c} & =-\frac{1}{4}\ln\left(1-g_{1}^{2}\right),\\
r_{d} & =-\frac{1}{4}\ln\left(1-g_{2}^{2}\right),
\end{align*}
resulting in the form 

\begin{align*}
S^{\dagger}\left[r_{c}\right]\mathcal{H}_{1}S\left[r_{c}\right] & =\omega_{0}\left(\sqrt{\left(1-g_{1}^{2}\right)}-1\right)c^{\dagger}c+E_{G_{1}}\\
S^{\dagger}\left[r_{d}\right]\mathcal{H}_{2}S\left[r_{d}\right] & =\omega_{0}\left(\sqrt{\left(1-g_{2}^{2}\right)}-1\right)d^{\dagger}d+E_{G_{2}}
\end{align*}

with the ground state energy 

\begin{align*}
E_{G_{1}} & =\frac{\omega_{0}\left(\sqrt{\left(1-g_{1}^{2}\right)}-1\right)}{2}-\frac{\omega_{0}}{4}J,\\
E_{G_{2}} & =\frac{\omega_{0}\left(\sqrt{\left(1-g_{1}^{2}\right)}-1\right)}{2}+\frac{\omega_{0}}{4}J.
\end{align*}

The eigenstates for $\mathcal{H}_{1}$ and $\mathcal{H}_{2}$ are
respectively

\begin{align*}
|\varphi_{1}^{n_{1}}\rangle & =S\left[r_{c}\right]|n_{1}\rangle,\\
|\varphi_{2}^{n_{2}}\rangle & =S\left[r_{d}\right]|n_{2}\rangle.
\end{align*}

\subsection*{Unsqueezed and squeezed STA driving}

Consider an arbitray Hamiltonian given by

\[
H\left(t\right)=\omega_{0}a^{\dagger}a-\frac{\omega_{0}}{4}g^{2}\left(a^{\dagger}+a\right)^{2},
\]
which the eigenstates are squeezed, and the Hamiltonian 

\[
\mathcal{H}\left(t\right)=\omega_{t}\left(b^{\dagger}b+1/2\right)-\frac{\omega_{0}}{2},
\]
with $\omega_{t}=\sqrt{\omega_{0}^{2}\left(1-g^{2}\left(t\right)\right)}$.
The relation between the modes are given by

\begin{align*}
b^{\dagger} & =\frac{1}{2}\left[\left(a^{\dagger}+a\right)a_{t}^{1/4}+\left(a^{\dagger}-a\right)a_{t}^{-1/4}\right]\\
b & =\frac{1}{2}\left[\left(a^{\dagger}+a\right)a_{t}^{1/4}-\left(a^{\dagger}-a\right)a_{t}^{-1/4}\right],
\end{align*}
 where we defined $a\left(t\right)=1-g^{2}\left(t\right)$. 

By applying the transformation on the operators in the STA driving
Hamiltonian for the unsqueezed harmonic oscillator, we obtain

\begin{align*}
b^{2}-b^{\dagger2} & =\frac{1}{4}\left[\left(a^{\dagger}+a\right)a_{t}^{1/4}-\left(a^{\dagger}-a\right)a_{t}^{-1/4}\right]^{2}\\
 & -\frac{1}{4}\left[\left(a^{\dagger}+a\right)a_{t}^{1/4}+\left(a^{\dagger}-a\right)a_{t}^{-1/4}\right]^{2}\\
 & =a^{2}-a^{\dagger2},
\end{align*}
 evidencing that the STA driving Hamiltonian is the same for both
cases.

\end{document}